\documentclass{article}

\usepackage{arxiv}

\usepackage{graphicx}
\usepackage{dcolumn}
\usepackage{bm}
\usepackage{multirow}
\usepackage{amsmath}
\usepackage{amssymb}
\usepackage{url}
\usepackage{pdfpages}

\date{\today}

\author{
Roberto Dale \\
Centro de Investigaci\'{o}n Operativa, \\
Universidad Miguel Hern\'{a}ndez, \\
Elche, 03202 Alicante, Spain \\
\texttt{rdale@umh.es}
\And
Ramon Lapiedra \\
Departament d'Astronomia i Astrof\'isica,  \\
Universitat de Val\`encia, \\
46100 Burjassot, Val\`encia, Spain \\
Observatori Astron\`omic, \\
Universitat de Val\`encia, \\
E-46980 Paterna, Val\`encia, Spain \\
\texttt{ramon.lapiedra@uv.es}
\And
Juan Antonio Morales-Lladosa \\
Departament d'Astronomia i Astrof\'isica,  \\
Universitat de Val\`encia, \\
46100 Burjassot, Val\`encia, Spain \\
Observatori Astron\`omic, \\
Universitat de Val\`encia, \\
E-46980 Paterna, Val\`encia, Spain \\
\texttt{antonio.morales@uv.es}
}

\begin{document}

\title{Cosmic primordial density fluctuations and Bell inequalities}
\maketitle

\begin{abstract}
The temperature measurements, $T$, of the perturbed cosmic microwave background,
performed by the cosmic background explorer satellite (COBE), are considered.
A dichotomist function, $f = \pm 1$,
is defined such that $f =+1$ if $\delta T > 0$ and $f =-1$  if $\delta T < 0$, 
where $\delta T$ stands for the fluctuation of $T$.
Then, it is assumed that behind the appearance of these fluctuations there is local realism.
Under this assumption, some specific Clauser-Horne-Shimony-Holt (CHSH) inequalities are
proved for these fluctuation temperatures measured by COBE in the different sky directions.
The calculation of these inequalities from the actual temperature measurements shows that these inequalities are not violated.
This result cannot be anticipated by calculating the commutators
of the cosmic density quantum operators. This must be remarked here since, in the case of a system of two
entangled spin ${\textstyle{1 \over 2}}$ particles, its CHSH inequalities violation can be inferred from the
nonvanishing value of the corresponding spin measurement commutators. The above nonviolation of the
observed cosmic CHSH inequalities is compatible with the existence of local realism behind the cosmic measurement results.
Nevertheless, assuming again local realism, some new cosmic CHSH inequalities can be derived for the case of the WMAP
measurements whose accuracy is better than the one of the above considered COBE measurements. More specifically,
in the WMAP case, some significant cross correlations between the temperature and polarization maps are detected,
and the new cosmic CHSH inequalities are the ones built with these cross correlations. Now, the occasional violation
of these CHSH inequalities would mean the failure of the assumed local realism in accordance with the quantum origin
of the primordial temperature and polarization fluctuations in the framework of standard inflation.

\end{abstract}

\tableofcontents

\section{Introduction}
\label{sec:1}

In the cosmic inflationary model, primordial density fluctuations could have a quantum mechanical origin \cite{Haw82,Gut82,Sta82,Bar83}.
Then, it would be very interesting to be able to test this kind of origin to rule out alternative classical scenarios \cite{Mal16},
as the scenarios considered in \cite{Ber95,Sen14}.

To begin with, notice that the universe produced by inflation is highly entangled.
Then, perhaps it could be possible to mimic the Bell inequalities and their possible violation \cite{Bel64} substituting the spin measurements by the corresponding cosmic density fluctuation measurements. However, there is a first difficulty to mimic Bell inequalities in this way such that the new inequalities can be violated. In order to explain the difficulty, we point out that the Bell-like inequalities we are going to consider here are some more general inequalities, named the Clauser-Horne-Shimony-Holt (CHSH) inequalities \cite{Cla69}, from the names of the authors. Then, we will introduce in the next section the expression of the CHSH inequalities, and we will show that these inequalities can only be violated if some commutators of the spin operators are nonzero, as can be seen by a simple inspection of the next identity, Eq.~(\ref{C_eq1}). On the other hand, substituting the present spin operators by the corresponding cosmic density operators, just mentioned before, and assuming local realism, we will obtain some, let us say, “cosmic CHSH inequalities.” Thus, to see whether these cosmic CHSH inequalities can be violated or not, and to conclude that they cannot, we will need first to clarify if some identity like Eq.~(\ref{C_eq1}) can be constructed from the corresponding cosmic density operators.

Now, in the beginning of this Introduction, we have talked of the inflationary origin of the primordial density fluctuations and implicitly of the corresponding measurements. Obviously, no observer was measuring anything at such a distant epoch, but, following the appearance of these fluctuations, they have been propagating across the cosmos reaching us now as the current observed spectrum of the cosmic microwave background. But, in this travel the primordial fluctuations become modified during the epoch of the thickness of the “last-scattering surface” (roughly speaking between the values of the redshift $z = 1100$ and $z = 1000$) due to different cosmic effects, called primary fluctuations (Sachs-Wolfe, adiabatic perturbations and Doppler effect) \cite{Sac67} and secondary ones added along the path between the last-scattering surface and today’s observer  (Rees-Sciama, gravitational lensing, Sunyaev-Zel’dovich, …) \cite{Ree68,Sel96,Sto99,Sun70}. Thus, even in the case that we could define some cosmic CHSH inequalities for this observed spectrum, it seems that we could not expect to find any violation of these hypothetical inequalities since the present observed fluctuations are not the same as the primordial ones, the only assumed to have a quantum origin. However, the degree of this modification of the primordial density fluctuations is very limited for large enough scales (though not, for shorter ones) \cite{Pee91,Pea98}. Then, perhaps we could expect a violation of the cosmic CHSH inequalities in the particular case of these large scales. Or in the alternative case where we could delimit the modification level of these primordial fluctuations. Nevertheless, in the present paper, we are not interested in these possibilities.

Our point of view is that, in the framework of the standard cosmology, after inflation, the Universe has evolved according to the laws of classical physics. Then, imagine that the appearance of the primordial density fluctuations at the end of inflation violates local realism. In such a case, the subsequent observed fluctuation spectrum of the cosmic microwave background, the fluctuation spectrum evolving according to the determinism of classical physics, will violate local realism too. This conclusion could be erroneous if, in their cosmic evolving, the primordial density fluctuations could lose their initial local character. But this is impossible, since, given any two primordial observing cosmic directions spatially separated in the relativistic sense of the word, this separated character increases in the cosmic time. Further, in our cosmic case, this primordial separation is present, roughly speaking, for $\alpha > 4$ arc degrees, where angle $\alpha$ is the one defined by both directions (see Appendix~\ref{App:1}), this limiting angle allowing the presence of data enough to make the corresponding statistical population.
Thus, our goal is to analyse whether the observed microwave background violates or not the corresponding cosmic CHSH inequalities. We will see that there is no such a violation in the present cosmic background explorer satellite (COBE)  \cite{Cob13} case and thus there is no failure of local realism in this case. In other words, as far as the primordial density fluctuations detected by the COBE measurements are concerned, there is no proof of a quantum origin of these fluctuations. If such a proof could be obtained, using WMAP or Planck measurement results \cite{WMAP17,Planck22} and violating some suitable cosmic CHSH inequalities, is an open question that we will consider at the end of the present paper.

The present paper is organised as follows. First, in Section~\ref{sec:2}, a well-known proof of the original CHSH inequalities \cite{Cla69} (for spin measurements) under the local realism hypothesis is given. In Section~\ref{sec:3}, a kind of CHSH inequality, involving cosmic density measurements instead of spin measurements, is also settled under local realism hypothesis. This kind of CHSH inequality is called “cosmic CHSH inequalities”. In Sections~\ref{sec:4} and~\ref{sec:5}, the calculation and computation results of the mean values appearing in these cosmic inequalities is presented. Then in Section~\ref{sec:6} it is shown that these cosmic CHSH inequalities are not violated, for the data of the COBE satellite \cite{Cob13}. Discussion in last section, including the possible violation of the new cosmic CHSH inequalities related to the WMAP or Planck measuring results, concludes the paper, before Appendix \ref{App:1}, dealing with the conditions to insure the cosmic angular spacelike separation of the measurements.


\section{Recalling the CHSH inequalities and their proof}
\label{sec:2}

Let us consider a quantum system of two entangled spin $\textstyle{1 \over 2}$ particles. The system has been prepared in such a way that the two particles move towards, respectively, two spin measurement devices placed at $A$ and $B$, the corresponding two measurement directions being $\vec x$ and $\vec y$. We will assume that the two measurement events are in a spacelike configuration, which means that we assume causality. After both measurements are performed a new run is considered: the system is prepared in the same state, the two measurements are performed again, and so on. Then, for each run, $\vec x$ takes randomly one of the two values, $\vec a$ or $\vec a'$, and similarly and independently, there are also two values $\vec b$ or $\vec b'$ for $\vec y$. Finally, depending on the context, $A$ and $B$, will represent too the corresponding measurement values. In this case we will have:
\begin{equation}  
A =  \pm 1,\quad B =  \pm 1,
\label{AB_val}
\end{equation}   
where the spin value has been normalised to unity.

Then, let us consider the four expecting values of the product $AB$ denoted by $\left\langle {AB} \right\rangle _{\vec x \vec y}$, according to the four possible cases
$(\vec a,\vec b),\,(\vec a,\vec b'),\,(\vec a',\vec b),\,(\vec a',\vec b')$.
With a view to facilitating the readability, we will be using a more compact notation: henceforth, $A$ and $A'$ will still represent the measurement operators in the place of $A$ in the directions $\vec a$ and $\vec a'$, respectively, and in the same way for $B$ and  $B'$. With this notation the four expecting values will be written
$\left\langle {AB} \right\rangle ,\,\left\langle {AB'} \right\rangle ,\,\left\langle {A'B} \right\rangle ,\,\left\langle {A'B'} \right\rangle$, with the values:
\begin{equation}  
A,\,A' =  \pm 1,\quad B,\,B' =  \pm 1.
\label{AApBBp_val}
\end{equation}   
Now let us define the operator $C$ as
\begin{equation}  
C \equiv A(B + B') + A'(B - B').
\label{C_def}
\end{equation} 
Assuming local realism, the following inequality, called CHSH inequality, can be proved:
\begin{equation}
\left |\left\langle C \right\rangle \right | = \left | \left\langle A(B + B') \right\rangle  + \left\langle A'(B - B') \right\rangle \right | \le 2.
\label{CHSH_ine}
\end{equation}

By definition, realism means that there exists some hidden variables, $\lambda$, 
previous to measurements, such that the values, $\pm 1$, of $A$, $A'$, $B$ and $B'$, are well-defined functions of
$\lambda$ and the corresponding measurement direction, that is, in an obvious notation, we are allowed to write
$A = A(\vec a,\lambda) =  \pm 1$ and $A' = A'(\vec a',\lambda) =  \pm 1$,
and similarly for $B$ and $B'$. As the used notation suggests, $A$ and $A'$ cannot depend on $\vec b$ or $\vec b'$
and the same for $B$ and $B'$ in relation to $\vec a$ and $\vec a'$. When this nondependence is assumed, as we do in the present case,
we say that the  assumed realism is local. This local character expresses the spacelike condition of the two measurement
events of each run, a condition that was demanded at the beginning of the present section.

Now, we follow the paper \cite{Mal16} to complete the proof of inequality (\ref{CHSH_ine}). First, notice that for each value of the
hidden variable we can have either $B = - B'$ or $B = B'$. But, because of the assumed causality, the measurement results for $B$ and $B'$
do not depend on whether we measure $A$ or $A'$. This means that in the two terms of the right-hand side of Eq.~(\ref{C_def}),
we must put the same equality, $B = - B'$ or, alternatively, $B = B'$, depending on the chosen $\lambda$ value.
In each of the two cases either the first term in Eq.~(\ref{C_def}) cancels or the last term cancels.
Therefore, the maximum value of $| C |$ is two. This completes the proof of inequality (\ref{CHSH_ine}).

To be more explicit, notice that, if the $B$ and $B'$ values were causally connected with the $A$ and $A'$ ones,
we should write something like $B_A$ and $B'_A$ in the first summand of the right-hand side of inequality (\ref{CHSH_ine}), 
and similarly, $B_{A'}$, $B'_{A'}$, in the second summand. Then, we had not been able to complete the proof.

To finish the present section, and following \cite{Mal16}, we take the square of the operator $C$. An elementary calculation gives:
\begin{equation}
{C^2} = 4 - [A,A'][B,B'],
\label{C_eq1}
\end{equation}
with the involution conditions $A^2 = A'^2 = I$ and $B^2 = B'^2 = I$ and where $[A,A']$ and $[B,B']$ stand for the corresponding commutators.
Obviously, only when both commutators are nonzero the inequality (\ref{CHSH_ine}), the CHSH inequality, may be violated.                                              

It is to be remarked that the inequality~(\ref{CHSH_ine}), on the one hand, and the expression~(\ref{C_eq1}), on the other hand, make reference to a very different kind of measurements. In~(\ref{CHSH_ine}), any of the four possible pairs of measurements, let us say $(A,B)$, $(A,B’)$, $(A’,B)$, or $(A’,B’)$, are performed on the same given state of the entangled system, and for simplicity, the two measurements of each pair can be taken at the same proper cosmic time. Differently, in~(\ref{C_eq1}), the four possible pairs of measurements, let us say $(A,A’)$, $(A’,A)$, $(B,B’)$, $(B’,B)$, are pairs of consecutive measurements, such that, the first one of any pair is performed on the same given initial state of the entangled system, while the second one is performed on the resulting collapsed state of the respective first one measured.

We will come back to the question in Section~\ref{sec:7}, where we show that in the cosmic case there is no an expression like~(\ref{C_eq1}) that would allow us to calculate indirectly the left-hand side of our cosmic CHSH inequality~(\ref{CHSH_ineR}). Contrarily to this, in the case of the entangled spin system described in the present section, the relation~(\ref{C_eq1}) can be  used to conclude indirectly if the corresponding CHSH inequalities~(\ref{CHSH_ine}) are or are not violated.

\section{The cosmic CHSH inequalities}
\label{sec:3}

Let us consider the cosmic microwave background with its anisotropic disturbance. Imagine that we measure its temperature, $T$, in many sky radial directions, $\vec x$, distributed in an isotropic way, like it has been done, for instance, by the COBE satellite. As we know, one finds
\begin{equation}
T (\vec x) = T_{0} + \delta T(\vec x),
\label{CMBT_eq1}
\end{equation}
with $T_{0}$ the mean uniform temperature, about $2.73^\circ K$, and $\delta T(\vec x)$ the depending on $\vec x$ disturbance.
Roughly speaking ${\delta T(\vec x) \mathord{\left/ {\vphantom {\delta T(\vec x) T_{0}}} \right. \kern-\nulldelimiterspace} T_{0}} \approx 5 \times 10^{-5}$
(see, for instance, \cite{Lid00}).

Now, let us consider the measuring values of $\delta T(\vec x)$ in the different directions, $\vec x$. Then, define the quantities $F(\vec x)$, in the following way:
\begin{subequations}
\label{Fx_def}
\begin{equation}
F(\vec x) = +1 \quad {\rm{if}} \quad \delta T(\vec x) > 0, \, {\rm{and}}
\end{equation}
\begin{equation}
F(\vec x) = -1 \quad {\rm{if}} \quad \delta T(\vec x) < 0.
\end{equation}
\end{subequations}

Further, we consider any two radial measurement directions, $\vec x$ and $\vec y$, whose common angle is $\alpha > 4^{\circ}$
arc degrees in order to ensure that the two sky places, identified by $\vec x$ and $\vec y$, are spacelike separated (see Appendix \ref{App:1}.
On the other hand, as explained next, at Section \ref{sec:41}, we should take $\alpha > 6^{\circ}$ according to Peebles~\cite{Pee91}. 

That is, if we denote by $\boldsymbol {\hat x}$ and $\boldsymbol {\hat y}$ the unit vectors along the outside directions $\vec x$ and $\vec y$, respectively, by definition we have
\begin{equation}
\cos\alpha \equiv {\boldsymbol  {\hat x}} \cdot {\boldsymbol  {\hat y}}.
\label{Cos_alpha}
\end{equation}

We will denote by $\left< F({\boldsymbol  {\hat x}})F({\boldsymbol  {\hat y}}),\alpha\right >$ the mean value of the product
$F({\boldsymbol  {\hat x}})F({\boldsymbol  {\hat y}})$ all over the direction pairs satisfying Eq.~(\ref{Cos_alpha}) for a given constant angle $\alpha$.
It is to be understood that, regarding these measurement pairs, the only ones to be considered are the ones present in the database of the measures actually performed.
Further, we will also use the simplified notation $\left < F({\boldsymbol  {\hat x}})F({\boldsymbol  {\hat y}}),\alpha\right > \equiv \left <\alpha\right >$.

Next we consider four different values for the constant angle $\alpha$, that is $\alpha_i$, $i = 1,2,3,4$, and the corresponding four mean values $\left < \alpha_{i} \right >$.
Now, the question is the following one: can we, using these cosmic four mean values, prove some cosmic inequalities similar to the proven CHSH inequalities  (\ref{CHSH_ine}),
under the local realism assumption? 

\subsection{Local realism: The cosmic case}
\label{sec:31}

Here, in this cosmic case, we define local realism in the same way we have already defined it in the precedent section (in the paragraph just after Eq.~(\ref{CHSH_ine})). Thus, we will say that we have local realism in our cosmic case if some hidden variables, $\lambda$, previous to the measurements, exist such that the following function, $f$, is defined:
\begin{equation}
f\left( {\boldsymbol {\hat x},\lambda } \right) = F\left( {\boldsymbol {\hat x}} \right) =  \pm 1,
\label{Fxl_def}
\end{equation}
with the condition that $f$ cannot depend on any direction $\boldsymbol {\hat y}$, whose sky image be spacelike separated from the image of $\boldsymbol {\hat x}$.

In the inflationary cosmic model, if we assume local realism, $\lambda$ must include the particular cosmic time, $t$, in which the end of inflation occurs at each sky locality pointed out by direction $\boldsymbol {\hat x}$. Then, in a more explicit notation, we will write Eq.~(\ref{Fxl_def}) as:
\begin{equation}
f\left( {\boldsymbol {\hat x},t,\lambda } \right) = F\left( {\boldsymbol {\hat x}} \right) =  \pm 1,
\label{Fxtl_def}
\end{equation}

In all, could we mimic, in the present cosmic case, the proof of the original CHSH inequalities, expression (\ref{CHSH_ine}), performed in the previous section, in order to prove some similar CHSH cosmic inequalities? 

\subsection{Mimicking the CHSH inequalities in the cosmic case}
\label{sec:32}

To begin with, this mimicry could only be possible if the above four angles, $\left < \alpha_{i} \right >$, admit the existence of four unit vectors,
$\boldsymbol {\hat a},\, \boldsymbol {\hat a'},\, \boldsymbol {\hat b},\, \boldsymbol {\hat b'}$, such that:
\begin{equation}
\label{Cos_alphai}
\left.\begin{aligned}
{\boldsymbol  {\hat a}} \cdot {\boldsymbol  {\hat b}} &= \cos \alpha_{1}, \, {\boldsymbol  {\hat a}} \cdot {\boldsymbol  {\hat b'}} = \cos \alpha_{2}, \\
{\boldsymbol  {\hat a'}} \cdot {\boldsymbol  {\hat b}} &= \cos \alpha_{3}, \, {\boldsymbol  {\hat a'}} \cdot {\boldsymbol  {\hat b'}} = \cos \alpha_{4}.
\end{aligned}\right.
\end{equation}

If this is the case, the corresponding directions, ${\vec a},\, {\vec a'},\, {\vec b},\, {\vec b'}$, defined by these four unit vectors, could play a similar role than the one played by the equally named directions when proving ordinary CHSH inequalities in Section \ref{sec:2}. Thus, we show next that Eqs.~(\ref{Cos_alphai}) can be proved for some $\alpha_i$ four values, but not for any four ones. In order to do this, we express the hypothetical four vectors, $\boldsymbol {\hat a},\, \boldsymbol {\hat a'},\, \boldsymbol {\hat b},\, \boldsymbol {\hat b'}$,
through their components referred to some orthogonal frame. Then, without loss of generality, we always can write:
\begin{subequations}
\label{abab_OrtFr}
\begin{equation}
\label{abab_OrtFr_a}
\boldsymbol {\hat a} = (1,0,0),
\end{equation}
\begin{equation}
\label{abab_OrtFr_b}
\boldsymbol {\hat a'} = (\cos \beta,\sin \beta,0),
\end{equation}
\begin{equation}
\label{abab_OrtFr_c}
\boldsymbol {\hat b} = (\cos \theta,\sin \theta \cos\rho,\sin \theta \sin \rho),
\end{equation}
\begin{equation}
\label{abab_OrtFr_d}
\boldsymbol {\hat b'} = (\cos \varphi,\sin \varphi \cos\mu,\sin \varphi \sin \mu),
\end{equation}
\end{subequations}
with angles $\beta, \, \theta,\,\varphi$ taking values between $0$ and $\pi$, and $\rho, \, \mu$, between $0$ and $2 \pi$.
Hence, the hypothetical four unit vectors $\boldsymbol {\hat a},\, \boldsymbol {\hat a'},\, \boldsymbol {\hat b},\, \boldsymbol {\hat b'}$,
solution of Eqs.~(\ref{Cos_alphai}), should satisfy the conditions
\begin{subequations}
\label{alphai_cond}
\begin{equation}
\label{alphai_cond_a}
\cos \theta = \cos \alpha_{1},
\end{equation}
\begin{equation}
\label{alphai_cond_b}
\cos \varphi = \cos \alpha_{2},
\end{equation}
\begin{equation}
\label{alphai_cond_c}
\cos \beta  \cos \alpha_{1} + \sin \beta  \sin \alpha_{1} \cos \rho = \cos \alpha_{3},
\end{equation}
\begin{equation}
\label{alphai_cond_d}
\cos \beta  \cos \alpha_{2} + \sin \beta  \sin \alpha_{2} \cos \mu = \cos \alpha_{4},
\end{equation}
\end{subequations}
for each value of $\beta$, that is, for each vector $\boldsymbol {\hat a'}$. Now, Eqs.~(\ref{alphai_cond_c}) and (\ref{alphai_cond_d}), can be written
\begin{subequations}
\label{rhomu_cond}
\begin{equation}
\label{rhomu_cond_a}
\cos \rho = \frac {\cos \alpha_{3} -\cos \beta \cos \alpha_{1}}{\sin \beta \sin \alpha_{1}},
\end{equation}
\begin{equation}
\label{rhomu_cond_b}
\cos \mu = \frac {\cos \alpha_{4} -\cos \beta \cos \alpha_{2}}{\sin \beta \sin \alpha_{2}}.
\end{equation}
\end{subequations}
But since $\left| \cos \rho \right | < 1$ and $\left| \cos \mu \right | < 1$, the following inequalities,
\begin{subequations}
\label{alphai_cond}
\begin{equation}
\label{alphai_cond_a}
\left |{\cos \alpha_{3} -\cos \beta \cos \alpha_{1}} \right | \le {\sin \beta \sin \alpha_{1}},
\end{equation}
\begin{equation}
\label{alphai_cond_b}
\left | {\cos \alpha_{4} -\cos \beta \cos \alpha_{2}} \right | \le {\sin \beta \sin \alpha_{2}},
\end{equation}
\end{subequations}
have to be satisfied. Later, in Section~\ref{sec:4}, we will prove that there exist some $\alpha_i$ values such that Eq.~(\ref{Cos_alphai}) are satisfied.
The proof is obtained by construction of the existing $\alpha_i$ through the suitable numerical calculations.

\subsection{Proof of the cosmic CHSH inequalities}
\label{sec:33}

\begin{figure*}[ht]
\begin{center}
\includegraphics[scale=0.58]{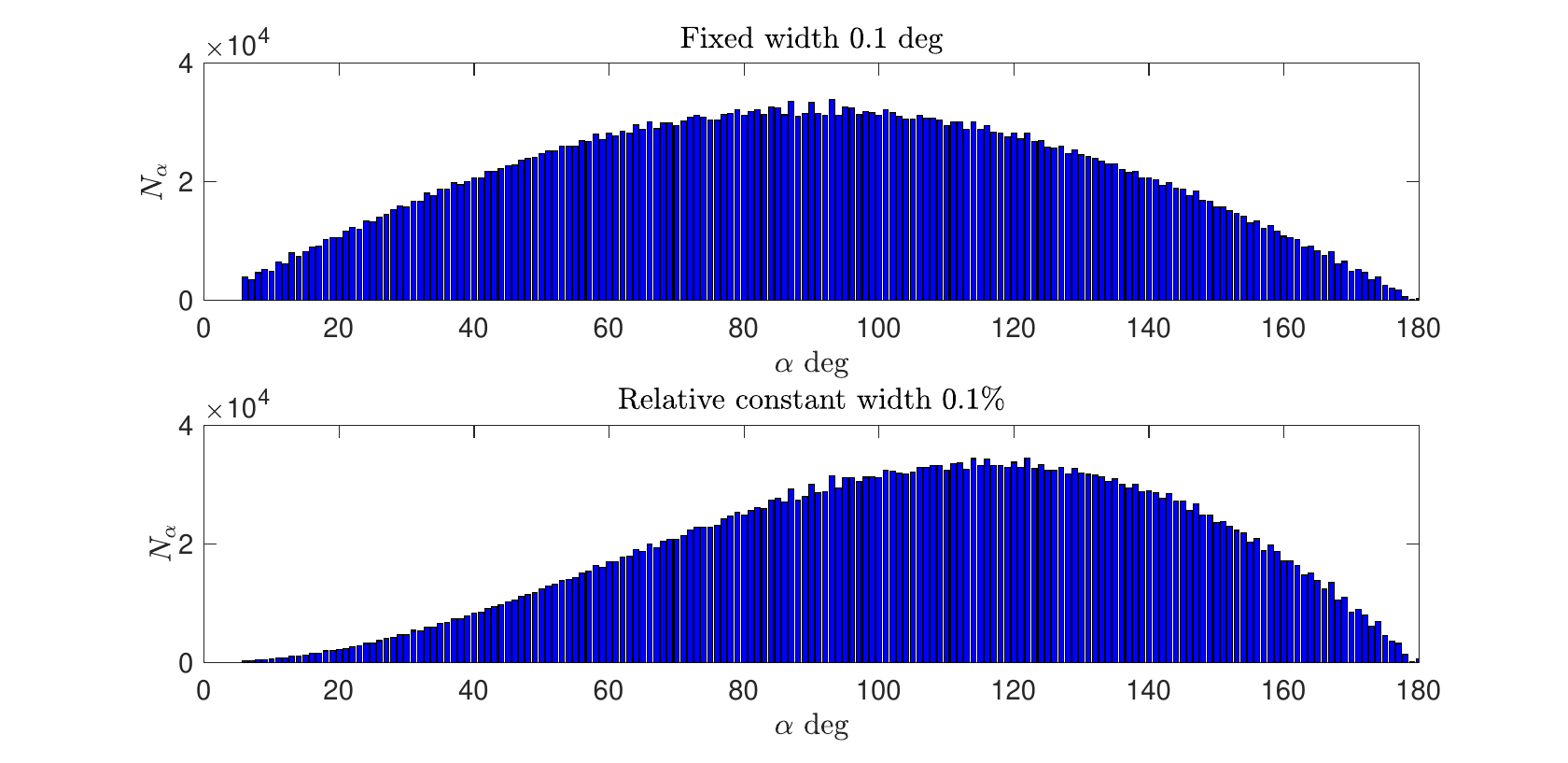}
\end{center}
\caption{The number of unit vector pairs, $N_\alpha$, contained in a certain interval centred at $\alpha$, that is
$\left[ {\alpha  - \delta \alpha ,\alpha  + \delta \alpha } \right]$, is presented. In the upper panel the interval width is built with a constant value of $\delta \alpha = 0.1$ deg.,
while the bottom one is constructed from a constant relative percentage ${{\delta \alpha } \mathord{\left/  {\vphantom {{\delta \alpha } \alpha }} \right. \kern-\nulldelimiterspace} \alpha } = 0.1\% .$}
\label{fig:Figure_01}
\end{figure*}

Then, since Eqs.~(\ref{Cos_alphai}) can be satisfied for some $\alpha_{i}$ values, it seems that all we need to prove some cosmic CHSH inequalities is to reproduce here,
for these $\alpha_{i}$, the argument developed after Eq.~(\ref{CHSH_ine}) which proves the  ordinary CHSH inequalities.
But, to verify that it is the case, we need to use a suitable version for the mean values, $\left< F({\boldsymbol  {\hat x}})F({\boldsymbol  {\hat y}}),\alpha\right >$,
$\boldsymbol {\hat x} = \boldsymbol {\hat a}, \, \boldsymbol {\hat a'}$, and $\boldsymbol {\hat y} = \boldsymbol {\hat b}, \, \boldsymbol {\hat b'}$,
appearing at the occasional cosmic CHSH inequalities. Actually, the number of summands, $n(\boldsymbol {\hat x},\boldsymbol {\hat y})$, involved in the mean value
$\left< F({\boldsymbol  {\hat x}})F({\boldsymbol  {\hat y}}),\alpha\right >$, depends, not exactly on the value of
$\cos\alpha \equiv {\boldsymbol  {\hat x}} \cdot {\boldsymbol  {\hat y}}$, but really of some narrow width $\delta \alpha$ centred on $\alpha$.
If we take this width, $\delta \alpha$, to be constant, i.e., independent of $\alpha$, then the corresponding number, $n(\boldsymbol {\hat x},\boldsymbol {\hat y})$,
depends on $\alpha$ (see Figure~\ref{fig:Figure_01} in Section~\ref{sec:42}).

Instead of this, the four expecting values, $\left <AB \right>, \, \left<AB' \right>$, etc., appearing at the ordinary CHSH inequalities (see the precedent section), involve the same number of summands. The mere consideration of this difference shows right away that the proof of the ordinary CHSH inequalities, such as they are, cannot be mimicked to produce a proof of the cosmic ones.

However, this difficulty can be easily circumvented by taking a variable $\delta \alpha$ width, that is, one depending on $\alpha$.
More precisely, a $\delta \alpha$ width fitted as to lead to a constant (that is, independent of $\alpha$) number $n(\boldsymbol {\hat x},\boldsymbol {\hat y})$.
Then, the door is open to mimic the proof of the ordinary CHSH inequalities performed in the Section \ref{sec:2},
for the purpose of proving  our cosmic CHSH inequalities. In all, assuming local realism, we want to prove the following cosmic CHSH inequalities
\begin{eqnarray}
\left.\begin{aligned}
\left | \left\langle C \right\rangle\right | &\equiv\left | \left\langle F({\boldsymbol  {\hat a}})F({\boldsymbol  {\hat b}}),\alpha_{1} \right\rangle
+ \left\langle F({\boldsymbol  {\hat a}})F({\boldsymbol  {\hat b'}}),\alpha_{2} \right\rangle \right.\\
&+ \left.\left\langle F({\boldsymbol  {\hat a'}})F({\boldsymbol  {\hat b}}),\alpha_{3} \right\rangle
- \left\langle F({\boldsymbol  {\hat a'}})F({\boldsymbol  {\hat b'}}),\alpha_{4} \right\rangle
\right | \le 2,
\label{CHSH_ine2}
\end{aligned}\right.
\end{eqnarray}
by mimicking the reasoning that in the previous section led to the original CHSH inequalities.

First of all, we will use a generic rotation, $R$, to write the mean values
$\left< F({\boldsymbol  {\hat x}})F({\boldsymbol  {\hat y}}),\alpha\right >$, of inequality (\ref{CHSH_ine2}),
in the alternative way $\left< F({R\boldsymbol  {\hat x}})F({R \boldsymbol  {\hat y}}) \right >$.
With this notation, the inequality~(\ref{CHSH_ine2}) can be written:
\begin{eqnarray}
\left.\begin{aligned}
\left | \left\langle C \right\rangle\right | &\equiv\left | \left\langle F(R{\boldsymbol  {\hat a}}) \left [ F(R{\boldsymbol  {\hat b}}) + F(R{\boldsymbol  {\hat b'}}) \right ] \right\rangle \right.\\
&+ \left. \left\langle F(R{\boldsymbol  {\hat a'}}) \left [ F(R{\boldsymbol  {\hat b}}) - F(R{\boldsymbol  {\hat b'}}) \right ] \right\rangle
\right | \le 2,
\label{CHSH_ineR}
\end{aligned}\right.
\end{eqnarray}
where the ($\alpha_i$), defined by~(\ref{Cos_alphai}), remain implicit. 
Then, we can have either $F(R{\boldsymbol  {\hat b}}) = - F(R{\boldsymbol  {\hat b'}})$ or $F(R{\boldsymbol  {\hat b}}) = F(R{\boldsymbol  {\hat b'}})$.
But, because of the assumed causality, the measurement results for the directions $R{\boldsymbol  {\hat b}}$ and $R{\boldsymbol  {\hat b'}}$
do not depend on whether we measure $R{\boldsymbol  {\hat a}}$ or $R{\boldsymbol  {\hat a'}}$.
This means that in the two terms of the right-hand side of inequality (\ref{CHSH_ineR}), we must put the same equality, 
$F(R{\boldsymbol  {\hat b}}) = - F(R{\boldsymbol  {\hat b'}})$ or $F(R{\boldsymbol  {\hat b}}) = F(R{\boldsymbol  {\hat b'}})$.
In each of the two cases either the first term in inequality~(\ref{CHSH_ineR}) cancels or the last term one cancels.
Therefore, the maximum value of $\left | \left\langle C \right\rangle\right |$ is two. This completes the proof of inequality (\ref{CHSH_ineR}).

\section{Data processing}
\label{sec:4}

In next sections we will describe some of the most relevant aspects related with the numerical calculations we have performed. To begin with,
say that our final aim is to present the numerical results obtained for the quantities
$\left< F({R\boldsymbol  {\hat x}})F({R \boldsymbol  {\hat y}}), \alpha \right > \equiv \left < \alpha \right >$, described in the prior section,
which are the necessary ingredients of the inequalities (\ref{CHSH_ineR}). For this purpose we will use some of the datasets provided by COBE satellite containing the proper necessary information, that, in this case, we can find in Ref.~\cite{Cob13}. Detailed information can be found, for instance, in \cite{Mat90,Mat94,Mat99}.

\subsection{On the data source}
\label{sec:41}

From the downloaded COBE FITS (flexible image transport system) image file, including maps of the CMB temperature and its uncertainty, certain parameters and other data, a table with the galactic coordinates (longitude, latitude) and its corresponding temperature (including the uncertainty) is built. This file contains a total of 6067 records which accounts for the same number of different possible vector directions in the sky. 

Next step is to design a set of unit vector pairs that will define, through the expression (\ref{Cos_alpha}),
a population of $0 < \alpha \le \pi$ angles. To achieve it, we combined all these 6067 directions obtaining a total number of 18,401,211 angles.
If $\left (\theta_{i},\varphi_{i} \right )$ and $\left (\theta_{j},\varphi_{j} \right )$ stand for the longitude and latitude coordinates of certain unit vector pair,
say $\boldsymbol {\hat x_{i}}$ and $\boldsymbol {\hat y_{j}}$, respectively, then the subtended angle $0 < \alpha_{ij} \le \pi$ is easily obtained from:
\begin{equation}
\label{alpha_ij}
{\alpha _{ij}} = \arccos \left [ {\cos {\varphi _i}\cos {\varphi _j}\cos \left( {{\theta _j} - {\theta _i}} \right) + \sin {\varphi _i}\sin {\varphi _j}} \right].
\end{equation}

But the available number of pairs has to be reduced discarding the causal connected directions at the last scattering surface. This can be achieved by imposing the condition
$\alpha_{ij} \gtrapprox 4^{\circ}$ (see Appendix~\ref{App:1} for details). However, according to Peebles (see \cite{Pee91}), during the galaxies formation epoch, the CMB is significantly altered due to the interaction with free electrons. Nevertheless this circumstance just affected small scales inhomogeneities, in such a way that $\alpha_{ij} > 6^{\circ}$ is a ''safety'' confidence margin. Taking values of $\alpha > 6^{\circ}$ do not suppose any loss of information in the numerical computation of the COBE dataset, whose angular resolution is of  $7^{\circ}$ \cite{Fix94}.

On the other hand, for each realization $\alpha_{ij}$ two temperatures are associated $T_{i}$ and $T_{j}$, linked with its corresponding directions
$\boldsymbol {\hat x_{i}}$ and $\boldsymbol {\hat y_{j}}$, respectively.

\subsection{Selection criteria}
\label{sec:42}

As mentioned in previous section, the calculation of each one of the quantities of interest $\left< F({\boldsymbol  {\hat x}})F({\boldsymbol  {\hat y}}),\alpha\right >$ for
a fixed value of $\alpha$, actually involves certain number of angles $\alpha_{ij}$, with $6^{\circ} < \alpha_{ij} \le 180^{\circ}$ included in a narrow interval, centred on $\alpha$, of width $\delta \alpha > 0$,
that is ${\alpha _{ij}} \in \left[ {\alpha  - \delta \alpha ,\alpha  + \delta \alpha } \right]$. In such a case, we have to specify the $\delta \alpha$ 
selection criteria and then, once it has been fixed, review the resulting sampling distribution. In Figure~\ref{fig:Figure_01} we represent the obtained results for a fixed value of
$\delta \alpha$ (same $\delta \alpha$ value for any of the considered $\alpha$ angle), and for a relative fixed value of $\delta \alpha$
(same ${{\delta \alpha } \mathord{\left/ {\vphantom {{\delta \alpha } \alpha }} \right. \kern-\nulldelimiterspace} \alpha }$ value for any of the considered $\alpha$ angle).

\begin{figure*}[ht]
\begin{center}
\includegraphics[scale=0.40]{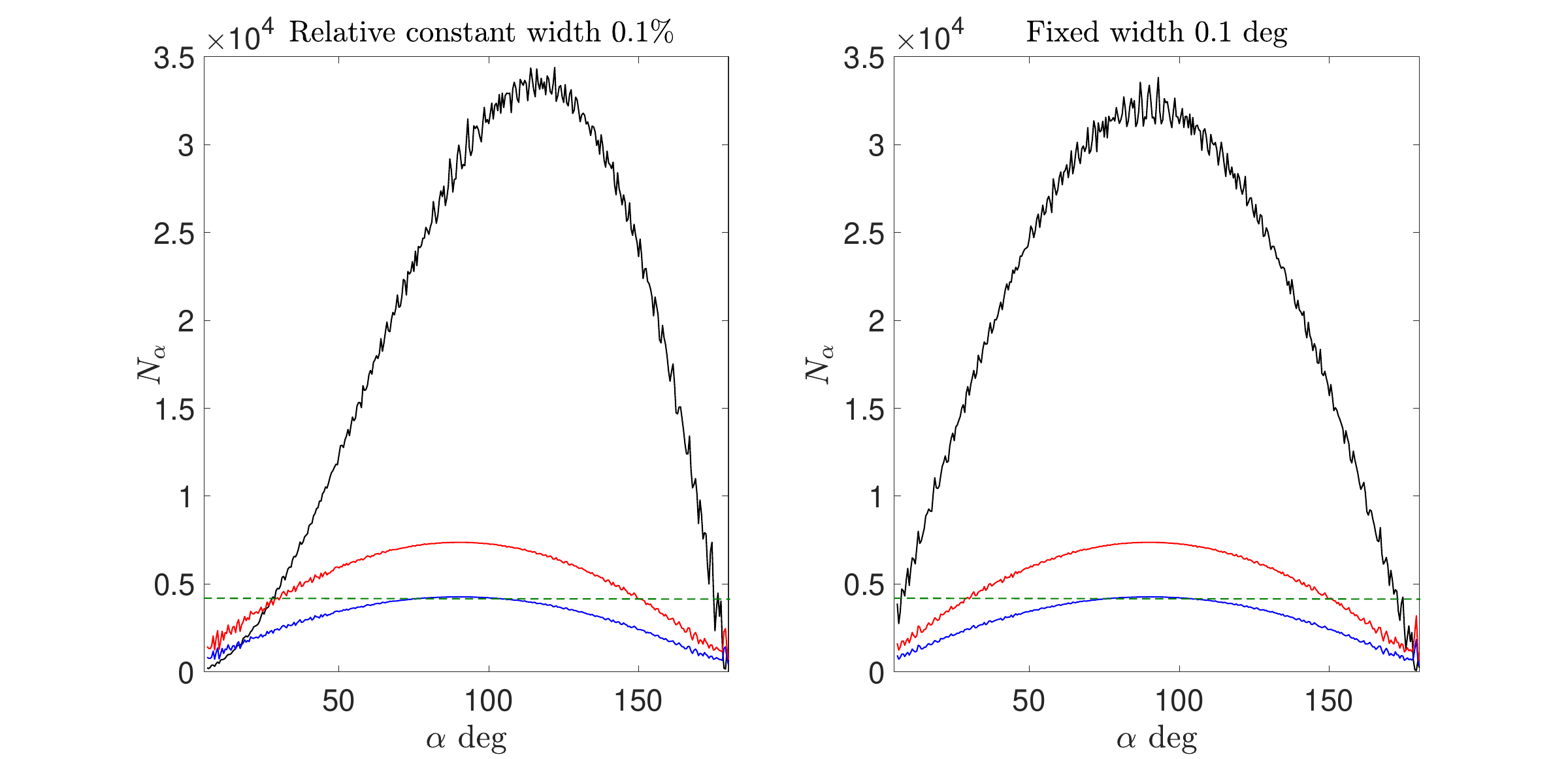}
\end{center}
\caption{These two illustration panels show several curves of a number of unit vector pairs, $N_\alpha$, as a function of $\alpha$, the left one for 
a constant relative percentage ${{\delta \alpha } \mathord{\left/  {\vphantom {{\delta \alpha } \alpha }} \right. \kern-\nulldelimiterspace} \alpha } = 0.1\%$, and
the right one when a constant value of $\delta \alpha = 0.1$ deg is considered. The solid blue curves (red solid curves) represent the minimum sample size for a confidence level
of $2\sigma$ ($3\sigma$), while the black one is the real amount of measures. Green slashed lines are the maximum value for the $2\sigma$ confidence level curves.}
\label{fig:Figure_02}
\end{figure*}

Different simulations have been performed, depending on the value of certain parameter values such as $\alpha$ stepping, $\alpha$ initial value or the interval width $\delta \alpha$.
As it can be appreciated in Figure~\ref{fig:Figure_01}, in any case, the number of vector pairs in each interval $\left[ {\alpha  - \delta \alpha ,\alpha  + \delta \alpha } \right]$,
named $n(\boldsymbol {\hat x},\boldsymbol {\hat y})$ in the previous section, varies with $\alpha$. Hence, the main component of our expecting value, that is a certain number of temperature pairs, depends on $\alpha$, and we may write $N_\alpha$.

\subsection{Sampling validation}
\label{sec:43}

Next we focus on the suitability of the sample size. To this end, statistical criteria have been used to establish a minimum population of items in each interval
$\left[ {\alpha  - \delta \alpha ,\alpha  + \delta \alpha } \right]$ \cite{Roh01}. In this sense, two confidence levels have been examined: $2\sigma \, (\sim 95\%)$ and
$3\sigma \, (\sim 99.7\%)$.

Now, remember that in Section~\ref{sec:33}, the constancy of $n(\boldsymbol {\hat x},\boldsymbol {\hat y})$ was required in order to be able to prove the expression~(\ref{CHSH_ineR}). Thus, for those confidence levels and for both cases considered in Figure~\ref{fig:Figure_01} (fixed $\delta \alpha$ or fixed
 ${{\delta \alpha } \mathord{\left/  {\vphantom {{\delta \alpha } \alpha }} \right. \kern-\nulldelimiterspace} \alpha }$),
the Figure~\ref{fig:Figure_02} shows that it is possible to select a constant value of $N_\alpha$
(see slashed green line, designed for the $2\sigma$ confidence level), in such a way that
\begin{itemize}
\item [(i)] It will be valid enough in terms of sample size, due to the fact that it has been selected as the maximum value of the $2\sigma$ curve (blue solid line).
\item [(ii)] The $\alpha$ covering range interval is maximised. The black curve represents the number of available pairs as a function of the $\alpha$ angle.
Observe that the only $\alpha$ range in degrees, delimited by the intersections of the aforementioned black curve and the slashed green line, that is excluded, is 
$\left] {0,12.5} \right[ \cup \left] {178,180} \right]$ for a relative width, and $\left] {0,29} \right[ \cup \left] {175,180} \right]$ for a fixed width.
\end{itemize}

According to both panels of Figure~\ref{fig:Figure_02}, it is explicit that inside of the nonexcluded range of the $\alpha$'s,
the number of pairs exceed the selected constant value of $N_\alpha$ (4,269 pairs in both cases).

Several tests have been executed: (i) the randomly selected over the oversampled pool have been repeated 50 times,
and (ii) the same calculations have been repeated again, including the full sample set for each $\alpha$. In this respect,
we will just add that, in the aforementioned pairs selection, the isotropy has been taken into account so that it covers the whole sky without any preferred direction.
The results have been about the same, actually the differences just affected the nonsignificant figures.

\section{Expected values present in the cosmic CHSH inequalities}
\label{sec:5}

\begin{figure*}[ht]
\begin{center}
\includegraphics[scale=0.58]{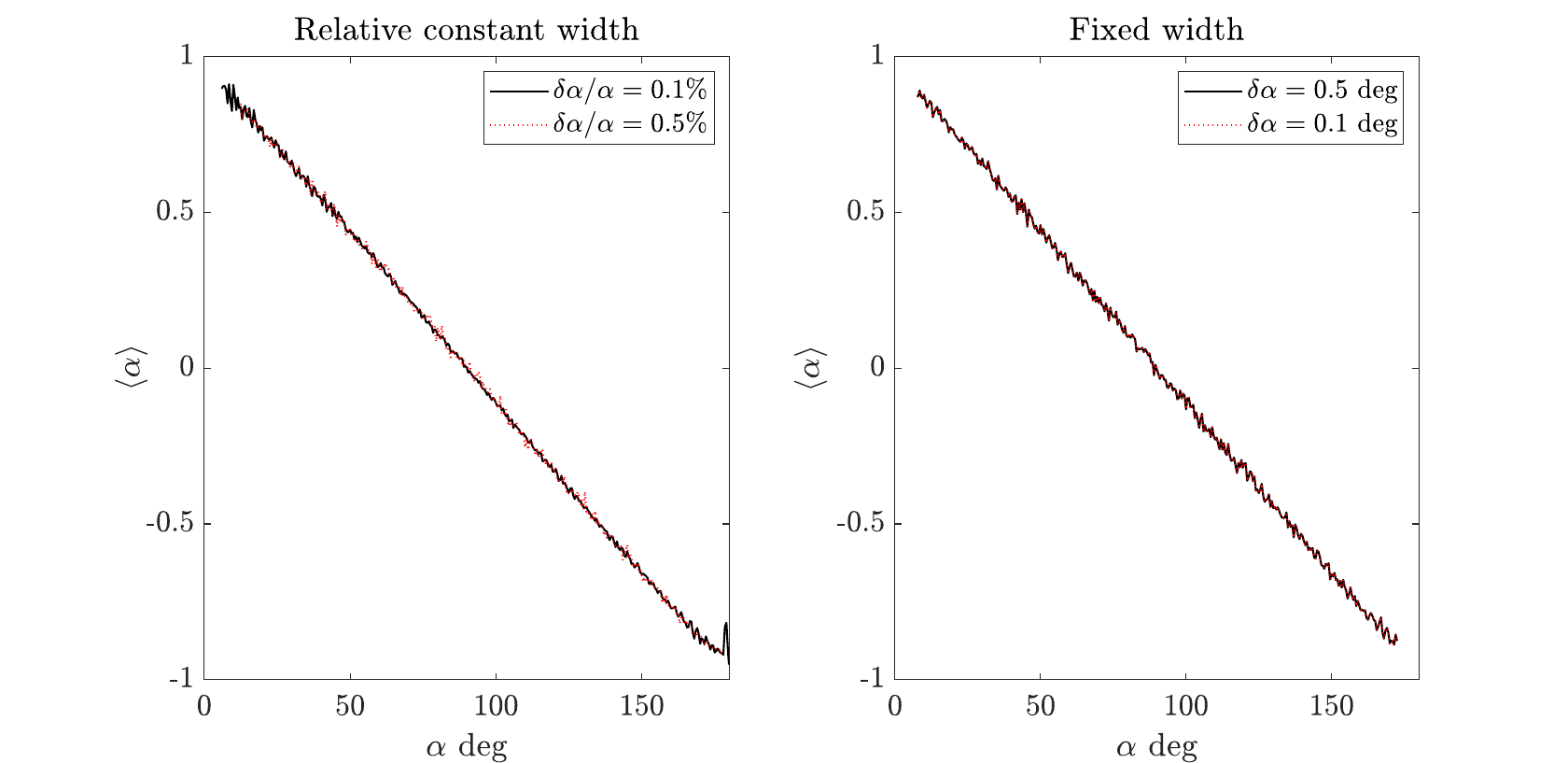}
\end{center}
\caption{Four different expected values curves  $\left\langle \alpha \right\rangle \equiv \left\langle {F\left( {\hat x} \right)F\left( {\hat y} \right),\alpha } \right\rangle$ 
are represented in these two panels. As in previous figures, two different types of intervals for $\alpha$ are shown: the left one corresponds to a
relative constant width while in the right one, a fixed width is settled. In both panels two cases are plotted:
${{\delta \alpha } \mathord{\left/  {\vphantom {{\delta \alpha } \alpha }} \right. \kern-\nulldelimiterspace} \alpha } = 0.1\%$ and
${{\delta \alpha } \mathord{\left/  {\vphantom {{\delta \alpha } \alpha }} \right. \kern-\nulldelimiterspace} \alpha } = 0.5\%$ for
the relative constant width case and ${\delta \alpha} = 0.5$ and ${\delta \alpha} = 0.1$ deg for the fixed width one.
For each case, the curve has been fitted to a straight line, see Table~\ref{tab:Table1}.}
\label{fig:Figure_03}
\end{figure*}
A set of selected data consists of a certain number of  $N_{\alpha}$ pairs of temperatures $\left( T_{i}, T_{j} \right)$ for a given value of
$\alpha_{ij}  \in \left[ {\alpha  - \delta \alpha ,\alpha  + \delta \alpha } \right]$. For each of a such selected dataset, the expected value
$\left< F({\boldsymbol  {\hat x}})F({\boldsymbol  {\hat y}}),\alpha\right >$ is calculated. This is accomplished through the following expression:

\begin{eqnarray}
\left.\begin{aligned}
\left\langle {F\left( {\boldsymbol  {\hat x}} \right)F\left( {\boldsymbol  {\hat y}} \right),\alpha } \right\rangle  &= P\left( { + , + ,\alpha } \right)  + P\left( { - , - ,\alpha } \right) \\ 
&- P\left( { + , - ,\alpha } \right) - P\left( { - , + ,\alpha } \right),
\label{Expected_V}
\end{aligned}\right.
\end{eqnarray}
where $P\left( {X,Y,\alpha } \right)$ stands for the probability of getting the result of $X\left( { \pm 1} \right)$ for $F({\boldsymbol  {\hat x}})$ and
$Y\left( { \pm 1} \right)$ for $F({\boldsymbol  {\hat y}})$ (see the $F$ function definition expressions~(\ref{Fx_def}) in Section~\ref{sec:3}),  for example
$P\left( { + , + ,\alpha } \right)$ represents the probability to obtain both temperatures of the two directions subtending the angle $\alpha$
greater than the mean uniform temperature $T_{0}$. The $T_{0}$ value that has been used is $T_{0}= 2.728 \pm 0.004$ K and it has been
selected from COBE files \cite{Cob13}. 

\subsection{Unmixed probabilities}
\label{sec:51}

Let us consider the accounting for the dichotomised value $\pm 1$ of function $F$ for certain direction $\boldsymbol  {\hat x}$, according to expressions~(\ref{Fx_def}) criteria. The frequency is established from $\delta T$ solely in such a way that its value is a $100\%$ $+1$ or $100\%$ $-1$. This is a first, easy and fast way of calculating. In the next subsection, a most sophisticated method for building probabilities will be developed.

Consequently, these probabilities in Eq.~(\ref{Expected_V}) are easily computed from

\begin{equation}
\label{P_alpha}
P\left( {X,Y,\alpha } \right) = \frac{{\nu \left( {X,Y,\alpha } \right)}}{N_{\alpha}},
\end{equation}
where $\nu\left( {X,Y,\alpha } \right)$ is the collected frequency for the resulting case $\left( {X,Y,\alpha } \right)$.
In Figure~\ref{fig:Figure_03} we have included, within two panels, some of the resulting curves of the function
$\left\langle {F\left( {\hat x} \right)F\left( {\hat y} \right),\alpha } \right\rangle$, wherein those ones analysed in
Figure~\ref{fig:Figure_01} and Figure~\ref{fig:Figure_02}, have also been incorporated.

Looking at the resulting curves enclosed in Figure~\ref{fig:Figure_03}, it is suggested that a suitable fit to a straight line,
$\left\langle \alpha  \right\rangle  = m\alpha  + n$, is possible. It is imperative to point out that, for the fit procedure,
just the validated interval range for the alpha angles – described in prior section – has been taken into account.
The results of those fittings, using the method of least squares, are summarised in Table~\ref{tab:Table1}.
From these results two immediate assertions can be done: (i) the ${{\cal R}^2} \sim 0.999$
coefficient of determination indicates a reliable fit, and (ii) the obtained values for fitting parameters $m$ and $n$ are almost equal, 
independently of the considered width, $m \approx 0.62$ and $n \approx -0.97$.

\begin{table*}
    \centering
    \begin{tabular}{|c|r|r|r|r|}
    \hline
    Width & \multicolumn{1}{c|}{$m \, (\alpha$ in deg)} & \multicolumn{1}{c|}{$m  \, (\alpha$ in rad)} & \multicolumn{1}{c|}{$n$}
    & \multicolumn{1}{c|}{${\cal R}^2$} \\ \hline
    $0.10\%$ & $-0.01090 \pm 0.00004$ & $-0.6245 \pm 0.0023$ & $0.981 \pm 0.004$ &  $0.9991$ \\
    $0.50\%$ & $-0.01088 \pm 0.00003$ & $-0.6234 \pm 0.0023$ & $0.980 \pm 0.004$ & $0.9992$ \\
    $0.05$ deg & $-0.01094 \pm 0.00002 $ & $-0.6234 \pm 0.0011$ & $0.9845 \pm 0.0012$ & $0.9991$ \\
    $0.10$ deg & $-0.01089\pm 0.00004$ & $-0.6240 \pm 0.0023$ & $0.979 \pm 0.004 $ & $0.9992$ \\ \hline
    \end{tabular}
  \caption{Information about some of our resulting fits to a straight line, obtained through the least squares method.
The first column indicates the selected $\delta \alpha$ width, columns two and three contain the obtained value for the slope of the line, $m$,  in degrees
or radians including the error, respectively. The forth one contains the intercept $n$ and its error, and the final one encloses the ${\cal R}^2$ goodness estimator.
In the first two rows, a relative constant width ${{\delta \alpha } \mathord{\left/  {\vphantom {{\delta \alpha } \alpha }} \right. \kern-\nulldelimiterspace} \alpha }$
is considered, while the third and fourth ones refer to the $\alpha$ fixed width case.}
  \label{tab:Table1}
\end{table*}

\subsection{Mixed probabilities}
\label{sec:52}

On this occasion our starting point will be the probabilities $p_{\pm} (\boldsymbol {\hat x})$ of getting the dichotomise value $\pm 1$ of function $F({\boldsymbol {\hat x}})$ defined in statements~(\ref{Fx_def}), respectively. These probabilities are built considering the absolute error in $\delta T(\boldsymbol {\hat x})$ quantities that are derived from the COBE original dataset. 

Hereafter, for a fixed direction $\boldsymbol {\hat x}$, the   $\varepsilon_{a}(\delta T)$ will represent the absolute error in $\delta T$. Based on the $\delta T$ interval defined by $[\delta T - \varepsilon_{a}(\delta T), \delta T + \varepsilon_{a}(\delta T) ]$ for a fixed direction $\boldsymbol {\hat x}$, the aforementioned $p_{\pm} (\boldsymbol {\hat x})$ are defined as follows: if we call $\varepsilon_{+}$ the positive section of the interval $[\delta T - \varepsilon_{a}(\delta T), \delta T + \varepsilon_{a}(\delta T) ]$ the probability $p_{+} (\boldsymbol {\hat x})$ can be defined as

\begin{equation}
\label{P_+}
p_{+}(\boldsymbol {\hat x}) \equiv \frac {\varepsilon_{+}} {2 \varepsilon_{a}(\delta T)},
\end{equation}
while
\begin{equation}
\label{P_-}
p_{-}(\boldsymbol {\hat x}) \equiv \frac {{2 \varepsilon_{a}(\delta T)} - \varepsilon_{+}} {2 \varepsilon_{a}(\delta T)},
\end{equation}
where, for simplicity, the $\boldsymbol {\hat x}$ dependence in $\delta T$ has been omitted.

From this definition, the four probabilities $p_{++}$, $p_{+-}$,  $p_{-+}$ and $p_{--}$ representing the probability of getting the values $++$, $+-$, $-+$ and $--$, respectively, for a single vector pair $(\boldsymbol {\hat x},\boldsymbol {\hat y})$ are $p_{XY}(\boldsymbol {\hat x},\boldsymbol {\hat y}) = p_{X}(\boldsymbol {\hat x}) \cdot p_{Y} (\boldsymbol {\hat y})$.

Finally, the probabilities $P(X,Y,\alpha)$ that appear in equation~(\ref{Expected_V}), assuming we have a total number of $N_{\alpha}$ pairs of vectors  $(\boldsymbol {\hat x},\boldsymbol {\hat y})$ subtending certain angle $\alpha$, are

\begin{equation}
\label{P_alpha_2}
P\left( {X,Y,\alpha } \right) = 
\sum\limits_{i=1}^{{N_\alpha }} {\frac{{{p_{XY}(\boldsymbol {\hat x},\boldsymbol {\hat y})_i}}}{{{N_\alpha }}}},
\end{equation}
with $X= \pm1 $, $Y= \pm 1$, and $N_\alpha$ the number of selected pairs of temperatures for each value of $\alpha$, following the criteria established in Section~\ref{sec:4}.
These statements allow us to calculate the expected value $\left< F({\boldsymbol {\hat x}})F({\boldsymbol {\hat y}}),\alpha\right >$ from expression~(\ref{Expected_V}).

For each angle $\alpha$, the absolute error, hereafter referred to as $\varepsilon_{XY}(\alpha)$, in the four $P(X,Y,\alpha)$ probabilities, is calculated at $2\sigma$ level thorough the standard statistics as \cite{Roh01}
\begin{equation}
\label{Error_P_alpha}
{\varepsilon _{XY}}\left( \alpha  \right) = Z\sqrt {\frac{{P\left( {X,Y,\alpha } \right)\left[ {1 - P\left( {X,Y,\alpha } \right)} \right]}}{{{N_\alpha }}}},
\end{equation}
where $Z$ stands for $z$ score whose value is $Z=1.96$ at the $2\sigma$ level. Once these four errors have been collected, the error for the expected value
$\left\langle \alpha \right\rangle \equiv \left\langle {F\left( \boldsymbol {\hat x} \right)F\left( \boldsymbol {\hat y} \right),\alpha } \right\rangle$ defined in~(\ref{Expected_V}) is easily reached from
\begin{equation}
\label{Error_VE_alpha}
{\varepsilon _a}\left[ \left\langle \alpha \right\rangle \right] = {\left( {{\varepsilon _{ +  + }}^2 + {\varepsilon _{ +  - }}^2 + {\varepsilon _{ -  + }}^2 + {\varepsilon _{ -  - }}^2} \right)^{{\textstyle{1 \over 2}}}}.
\end{equation}
And the same technique provides the error for $\left| {\left\langle C \right\rangle } \right|$ quantity defined in~(\ref{CHSH_ine2}), built from four $\alpha$ values, 
say $\alpha_1$, $\alpha_2$, $\alpha_3$, $\alpha_4$; that is,

\begin{eqnarray}
\left.\begin{aligned}
{\varepsilon _a}\left( {\left| {\left\langle C \right\rangle } \right|} \right)  &= \left [ {\varepsilon _a}^2\left( {\left\langle {{\alpha _1}} \right\rangle } \right) +
{\varepsilon _a}^2\left( {\left\langle {{\alpha _2}} \right\rangle } \right) + {\varepsilon _a}^2\left( {\left\langle {{\alpha _3}} \right\rangle } \right) \right.\\
&+ \left. {\varepsilon _a}^2\left( {\left\langle {{\alpha _4}} \right\rangle } \right)
\right ]^{{\textstyle{1 \over 2}}}.
\label{Error_C}
\end{aligned}\right.
\end{eqnarray}

\begin{figure*}[ht]
\begin{center}
\includegraphics[scale=0.58]{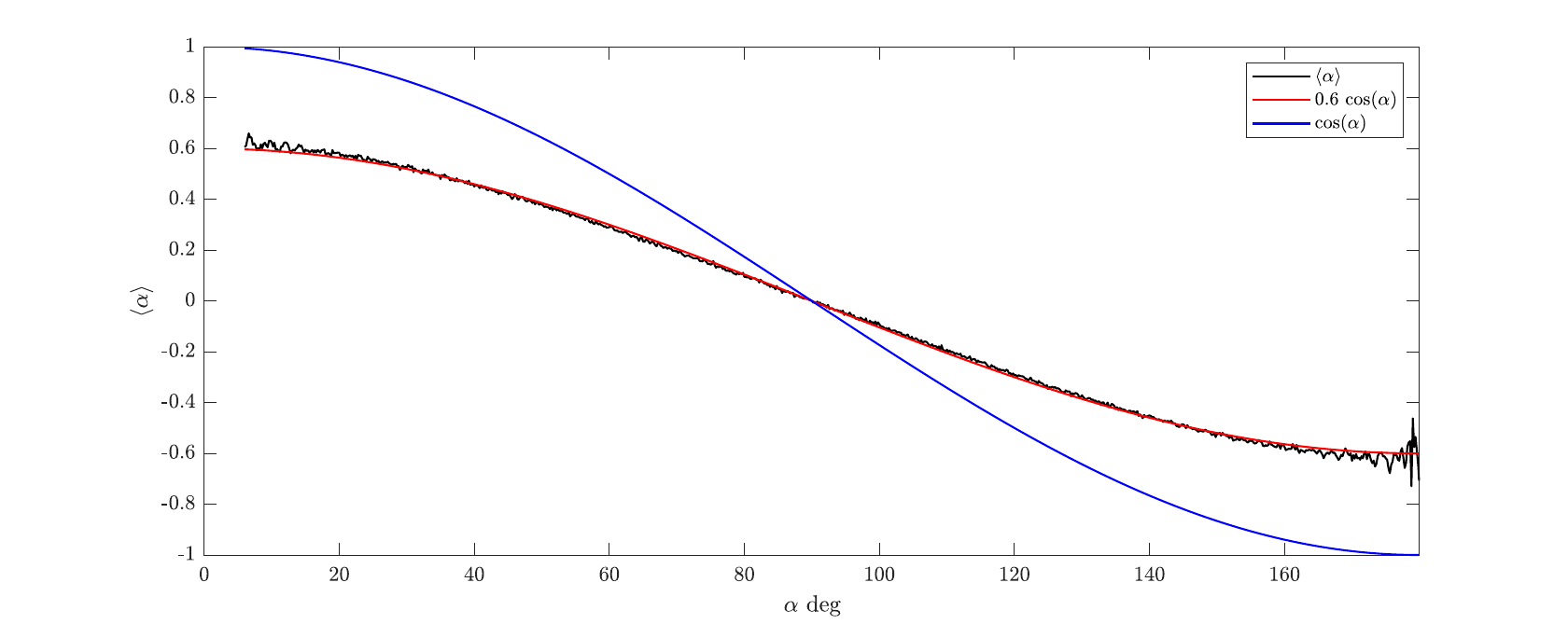}
\end{center}
\caption{The expected values curve  $\left\langle \alpha \right\rangle \equiv \left\langle {F\left( {\hat x} \right)F\left( {\hat y} \right),\alpha } \right\rangle$, 
calculated through a mixed probabilities method, is presented in this figure for a fixed width $\delta \alpha=0.05$ deg.
The obtained curve $\left\langle \alpha \right\rangle$ is well approached by the function $0.6 \, \cos \left( \alpha \right)$ (solid red line), while
the solid blue one is the function $\cos \left( \alpha \right)$.}

\label{fig:Figure_04}
\end{figure*}

Using this method, a reasonable fit to a straight line, was not possible. An expected value  $\left\langle {F\left( {\hat x} \right)F\left( {\hat y} \right),\alpha } \right\rangle$ sample curve is plotted in Figure~\ref{fig:Figure_04}, it has been built accurately with a small step value for $\alpha$ (0.2 deg) and a narrow width $\delta \alpha$ (0.05 deg).
Notice that, this expected value curve, that results when the mixed probabilities are considered, is some more similar to a cosine type curve. Both curves are plotted together in Fig.~\ref{fig:Figure_04}. Remarkably, the obtained expected value curve is a perfectly weighted cosine function, having an amplitude of 0.6, that is $\left\langle \alpha  \right\rangle \left( \alpha  \right) \approx 0.6\cos \left( \alpha  \right)$.

\section{Numerical evaluation of the $\left | \left\langle C \right\rangle\right |$ values and the nonviolation of the cosmic CHSH inequalities}
\label{sec:6}
Next, to test the cosmic CHSH inequalities, let us return to the main expression~(\ref{CHSH_ineR}) and provide some details about the followed procedure to test the aforementioned inequality:
\begin{itemize}
\item [(i)] For all the different feasible expected values analysed, a scanning code procedure, whose objective of finding four $\alpha$’s values which maximised function $\left | \left\langle C \right\rangle\right |$, was written.
\item [(ii)] Depending on the $\alpha$’s angles partition method (which defines a set of $\alpha_i$’s angles pool), a certain number of possible four angles combinations are obtained, those are our four-angle candidates. Nevertheless, not all these four-set candidates are viable, to be acceptable, conditions~(\ref {alphai_cond_a}) and~(\ref {alphai_cond_b}) have to be fulfilled. Once a four-set candidate has been selected, the only free parameter in the aforementioned conditions is the $\beta$ angle. So, a $\beta$ value that honours the referred conditions has to be found out. To accomplish this, a loop routine looks for it. If no $\beta$ value is found, the four-set candidate is discarded, on the contrary, if a $\beta$ value is found, the four-angle set is approved to be processed.
\item [(iii)]For each acceptable fourset, the $\left | \left\langle C \right\rangle\right |$ is computed. Those calculated $\left | \left\langle C \right\rangle\right |$ values that exceed a preestablished threshold (and its corresponding four $\alpha$'s) are stored into a file and then, this resulting file is analysed.
\end{itemize}

For instance, for a partition of 871 $\alpha$’s angles there are a number of $23,815,827,035$ combinations, however there are four permutations of  $\left\{ {{\alpha _1},{\alpha _2},{\alpha _3},{\alpha _4}} \right\}$ which leave $\left | \left\langle C \right\rangle\right |$ unaltered. In such a case, the total rises to $95,263,308,140$. As a result, a maximum value for $\left | \left\langle C \right\rangle\right |=1.71 \pm 0.03$ was achieved when $\left\{ {{\alpha _1}=125,{\alpha _2}=39.2,{\alpha _3}=135,{\alpha _4}=139} \right\}$ deg.

To conclude the current section, our simulations, that have been executed with different parameterisations (as $\alpha$ stepping, $\delta \alpha$ amplitude, etc.),
have confirmed that there is no cosmic CHSH inequalities~(\ref{CHSH_ine2}) violation, when COBE's data are considered. 

\section{Discussion and conclusions}   
\label{sec:7}

The main result of the previous section and of the all present paper is that the COBE measuring detections do not violate the cosmic CHSH inequalities~(\ref{CHSH_ineR}). A first comment to make about this result is that this no-violation cannot be inferred from a kind of relation such as~(\ref{C_eq1}) applied to our cosmic scenario, and so we have needed to calculate directly this no-violation. This is at variance with what happens in the case of the entangled spin system considered in Section~\ref{sec:2} where the no-violation of inequality~(\ref{CHSH_ine}) can be inferred from~(\ref{C_eq1}).

At first sight, it seems that in our cosmic case we could write an equation similar to~(\ref{C_eq1}), given that both cases, the cosmic one and the one referred to~(\ref{C_eq1}), respectively, involve the same sort of measurements, $A$, $A'$, $B$ and $B'$. However, the pairs of measurements actually performed from these four possible single measurements are very different according to the case, the cosmic one or the one related to~(\ref{C_eq1}), that we are considering. Thus, in this second case, let it be in particular the commutator $[A, A']$, present in the Eq.~(\ref{C_eq1}), or more specifically the product $A' A$. As explained at the end of the Sec.~\ref{sec:2}, this product describes the $A$ measurement performed on the initial state of the entangled spin system, followed by the $A'$ measurement on the collapsed state resulting of the $A$ measurement. But nothing similar to these two consecutive measurements can be present in our cosmic case. In this case, $A$ and $A’$ are two cosmic measurements performed, respectively, in two known sky directions, at two corresponding times belonging to the epoch of the end of inflation. But these two times are not known. Then, in particular, we do not know if they are or are not two consecutive times, which means that from the observational data we cannot infer the value of the corresponding $[A, A']$ commutator in the cosmic case. In other words, there is not a similar version of Eq.~(\ref{C_eq1}) in our cosmic case. There is not a similar version in the sense that we cannot use the commutator values to make easier the calculation of~(\ref{CHSH_ineR}) since these values remain unknown.  All the same, what is important for us is if the new cosmic CHSH inequality~(\ref{CHSH_ineR}) is violated or not, such that, looking for a similar cosmic equation to Eq.~(\ref{C_eq1}), a similar equation, actually nonexistent, would only have the interest of making the calculation of~(\ref{CHSH_ineR}) easier.

Therefore, it makes sense to test the validity or violation of the cosmic CHSH inequalities, and to quantify the results. The mixed probabilities treatment (used in Section~\ref{sec:52}) provides an expected values curve $\left\langle \alpha \right\rangle = 0.6 \, \cos(\alpha)$. This $\cos(\alpha)$ shape is the expectation value for the product of two spin $1 \over 2$ measurements in the directions of two unit vectors $\vec a$ and $\vec b$, on a suitable entangled quantum system of two spin $1 \over 2$ particles, such that $\vec a \cdot \vec b = \cos(\alpha)$. See Ref.~\cite{Bel64} for details.

To finish the paper, we could ask what would be the final result if, instead of working with the COBE data, we worked with the more accurate WMAP data~\cite{WMAP17} and Planck data~\cite{Planck22} which include polarization results. Would we still be able to prove, under the local realism assumption, a new cosmic CHSH inequality similar to~(\ref{CHSH_ineR})? Furthermore, a new cosmic inequality that became violated?

Trying to answer these questions, let us consider the following Kogut paper~\cite{Kog03}, where the WMAP has mapped the full sky in Stokes parameters $I$, $Q$, and $U$, the $I$ parameter corresponding to the temperature and $Q$ and $U$ parameters to the polarization. In Kogut's paper some significant correlations between the measured temperature and the measured polarization are detected. The simplest measure of this correlation is given by the two-point angular cross-correlation function (see this specific cross-correlation in the cited paper~\cite{Kog03}, equation (6)). In all, a correlation between $I$, on one hand, and $Q$, $U$, on the other hand. Then, on the base of this specific cross-correlation, a new cosmic CHSH inequality like Eq.~(\ref{CHSH_ineR}) can be proved under the local realism assumption (work in progress to be addressed elsewhere). More specifically, we can define
\begin{eqnarray}
\left.\begin{aligned}
\left\langle C_{IQ} \right\rangle & \equiv \left\langle F_I({\mathbf{\hat a}}) \, F_Q({\mathbf{\hat b}}),\alpha_{1} \right\rangle
+ \left\langle F_I({\mathbf{\hat a}}) \, F_Q({\mathbf{\hat b'}}),\alpha_{2} \right\rangle \\
&+ \left\langle F_I({\mathbf{\hat a'}}) \, F_Q({\mathbf{\hat b}}),\alpha_{3} \right\rangle
- \left\langle F_I({\mathbf{\hat a'}}) \, F_Q({\mathbf{\hat b'}}),\alpha_{4} \right\rangle,
\label{C_IQ_Def}
\end{aligned}\right.
\end{eqnarray}
where $F_I$ and  $F_Q$ are dichotomised measures for the corresponding Stokes parameters. Then, we can prove a sort of inequality~(\ref{CHSH_ineR}):
\begin{equation}
|\left\langle C_{IQ} \right\rangle| \leq 2\,  \epsilon , 
\label{Stokes_ine1}
\end{equation}

where the $I$ Stokes parameter is normalised to the unity and then dichotomised, such that it takes for each measure one of the two possible values $\pm 1$.
Furthermore, $\epsilon \approx 10^{-2}$ is the maximum value of the correction polarization terms to the above normalised and dichotomised $I$ parameter. Notice that this normalization and dichotomy are already present in the cosmic CHSH inequality~(\ref{CHSH_ineR}) that has been proved for the COBE data under the local realism assumption. Further, the inequality is in accordance with the resulting calculation from the data, as it has been shown in the preeding section.

This maximum $\epsilon$ value, $\epsilon \approx 10^{-2}$,  is a rough estimation that comes from the interpretation of Fig. 1 in Ref.~\cite{Kog03}, where $IQ$ correlations (in square micro-Kelvin) are given in terms of the angular separation $\theta$ (in degrees). This graph presents an absolute minimum  at the point $P \approx (35^{\circ}, - 7\mu K^{2}$). Thus, taking $\frac{\delta T(\vec{x})}{T_0} \approx 5 \times 10^{-5}$ for the mean values of relative disturbances in temperature intensities, the $\epsilon$ value at the point $P$ of the graphic is $\epsilon (P) \approx \frac{7}{25} \times 10^{-2}$.

But, the question is if we could expect that the new cosmic CHSH inequality, similar to~(\ref{CHSH_ineR}), that we have just commented on, would be violated for the WMAP or Planck data with their including polarization results. Our pedestrian but practical answer to the question is that, strictly speaking, we do not know any reasons to expect or not such a violation, but considering the great significance of this dilemma, that is, reject or not the local realism, and then reject or not the quantum origin of the inflation, we could find interesting to test in a future paper the possible violation, for the WMAP or Planck data, of our new cosmic CHSH inequalities~(\ref{Stokes_ine1}). Even more, in the case of cosmic CHSH inequalities~(\ref{CHSH_ineR}), where there is no violation, the measured quantity is a three scalar, the cosmic energy density, while in the case of an entangled spin system, where violation is present, the measured quantity is the spin. Finally, the measured quantities for the new cosmic CHSH inequalities, the WMAP/Planck data including the polarization results, involve three-vector directions. Then, the suspicion would be that, from the three cases we have just considered, only the one measuring three-scalar quantities would lead to a no violation of the corresponding CHSH inequalities, while in the new cosmic case (like happens too in the spin one) we could expect the presence of a violation. Then, it could be interesting to test such a violation.

We have studied a particular application to CMB, based on the original Bell inequality formulated for a two-partite system, and related with the two-point correlation function $C(\theta)$ of CMB temperature fluctuations (also with the equivalent power spectrum $C_\ell$). However, there exist extended inequalities for the multipartite case \cite{Mer90,Bel93}. A future task would be obtaining a new set of multipartitelike inequalities to be applied to the CMB. Deviations from the Gaussian character of the CMB spectrum are constrained from correlation measurement in temperature fluctuations on three and four sky positions. If such a deviation has a quantum origin, it should be manifested by extending the present study to check  a violation of these cosmic multipartite inequalities.

\section{Acknowledgements}
This work has been supported by the Spanish Ministerio
de Ciencia, Innovaci\'on y Universidades, Projects PID2019-109753GB-C21
and PID2019-109753GB-C22. We thank our colleagues \'Angel S\'anchez and Roland Calvo, from the statistics department of Universidad Miguel Hern\'andez, for their instructive comments.

\appendix

\section{THE NONCAUSAL CORRELATION CONDITION}
\label{App:1}

How far away do two of our cosmic measurement events, let us say $A$ and $B$, have to be to be able to ensure that they are in a spacelike configuration? To answer this question, we consider the comoving distance, $x_d$, travelled by a photon that leaves the \textit{observer} at the end of inflation and reaches its destination at the decoupling time (this is a purely geometric question, because we do not consider the incessant dispersion produced by an ionised medium). The differential equation for this photon trajectory $x(t)$ (in a spatially flat universe), taking the speed of light $c=1$, is
\begin{equation}\label{dx1}
d x = \frac{dt}{a} = \frac{d a}{a\dot{a}} = \frac{d a}{H a^2},				
\end{equation}
where $t$ refers to cosmological time, $a(t)$ is the cosmic scale factor, $H$ is the Hubble function, and  $\dot{a} \equiv \frac{d a}{dt}$.

On the other hand, setting $a_0 = 1$ at the present time,  the cosmic redshift $z$ is given by $1+z = a^{-1}$. Then 
$d z = - a^{-2} d a$,  and substituting in (\ref{dx1}), one has
\begin{equation}\label{dx2}
d x  =  - H^{-1} dz = - H_0^{-1} (1+z)^{-3/2} dz,	
\end{equation}
under the assumption of a matter dominant era (neglecting the cosmological constant contribution), with $H_{0}$ the Hubble constant.

Then, we will integrate Eq.~(\ref{dx2}) from the end of inflation, in a given locality, to, let us say, the decoupling instant. That is, in terms of $z$, from $z_{e} + \delta z(x)$ to $z_{d}$. Here, $z_{e}$, stands for the mean value of $z$ at the end of inflation, $\delta z(x)$ is the perturbation $of z_{e}$ in the cosmic locality $x$, and $z_{d}$ the value of $z$ for the decoupling instant. Hence, the corresponding integration of Eq.~(\ref{dx2}) gives for $\Delta x \equiv x_{d}$:
\begin{equation}\label{xd}
x_d = 2 H_0^{-1} \left [ (1+z_d)^{-1/2} - (1+ z_e +\delta z)^{-1/2}\right ],
\end{equation}
where $z_d \approx 1000$ and $z_e \approx 10^{26}$. Thus, since $z_e$ is huge compared to $z_d$, we have, irrespective of the value of $\delta z(x)$,
\begin{equation}\label{xd2}
x_d \approx 2 H_0^{-1} z_d^{-1/2}.
\end{equation}

Now, the above comoving distance value, $x_d$, has to be compared with the comoving last scattering surface diameter, $2 x_r$. To obtain $x_r$, let us take Eq.~(\ref{dx1}) with the opposited sign: $dx = - a^{-1} d t$. Thus, using $z$ for the referring cosmic time, one obtains Eq. (\ref{dx2}) with the reversed sign, which after integration from $z_d$ to $z = 0$, leads to
\begin{equation}\label{x-radio}				
x_r \approx - 2 H_0^{-1}.
\end{equation}
Therefore, we have the quotient value
\begin{equation}\label{ratio}
\frac{x_d}{|x_r|} = z_d^{-1/2} \approx (1100)^{-1/2} \approx \frac{1}{33},
\end{equation}
that is, to ensure that the measurement events $A$ and $B$ define a spacelike configuration, the angular separation between them has to be double the angle $\theta$ obtained from $\tan \left( \theta  \right) = \frac{{{x_d}}}{{\left| {{x_r}} \right|}}$, so $3^{\circ} \, 28'$ at least. 

To estimate the contribution of the cosmological constant, $\Lambda$,  in the previous reasoning, we should replace the relation:
\begin{equation}
\label{ambLambda} 
H = H_0 \sqrt{\Omega_m (1+z)^3 + \Omega_\Lambda}
\end{equation}
into (\ref{dx2}) and integrate the corresponding expressions numerically. Taking $\Omega_m = 0.3$, and $\Omega_\Lambda = 0.7$ for the matter density and cosmological constant parameters, respectively,  we obtain 
$\frac{x_d}{|x_r|} = \frac{1}{29}$, that is an angle $ \theta \approx 2^\circ$, which corresponds to an angular separation of $4^\circ$ at least for two measurement events $A$ and $B$ in spacelike configuration.


\newpage


\section*{Supplementary material}

# Addendum

After completing our analysis of the COBE FIRAS dataset, we are interested in extending our research using more advanced datasets from the WMAP and Planck satellites. It is worth noting that while COBE FIRAS had an angular resolution of 7 degrees, WMAP and Planck significantly improved this to 0.3 degrees and 5-10 arcminutes, respectively. Additionally, their accuracy has also been enhanced. The WMAP and Planck satellites measure the differences in temperature of the CMB between two different points in the sky (in opposite directions) rather than the absolute temperature values themselves as COBE FIRAS did (see [18]).

In this sense and with the aim of sharing interesting new results, a brief summary of our preliminary results is presented below in advance.

Using the techniques described in Subsection 5.2, we recalculated the quantities $\langle F(\hat{x})F(\hat{y}), \alpha \rangle$, as defined in (19). In the next two panel figure, the resulting curves of the mentioned function are shown. The left one has been built using the intensity values of the WMAP Q band data while the right one was produced using Planck NILC pipeline information.

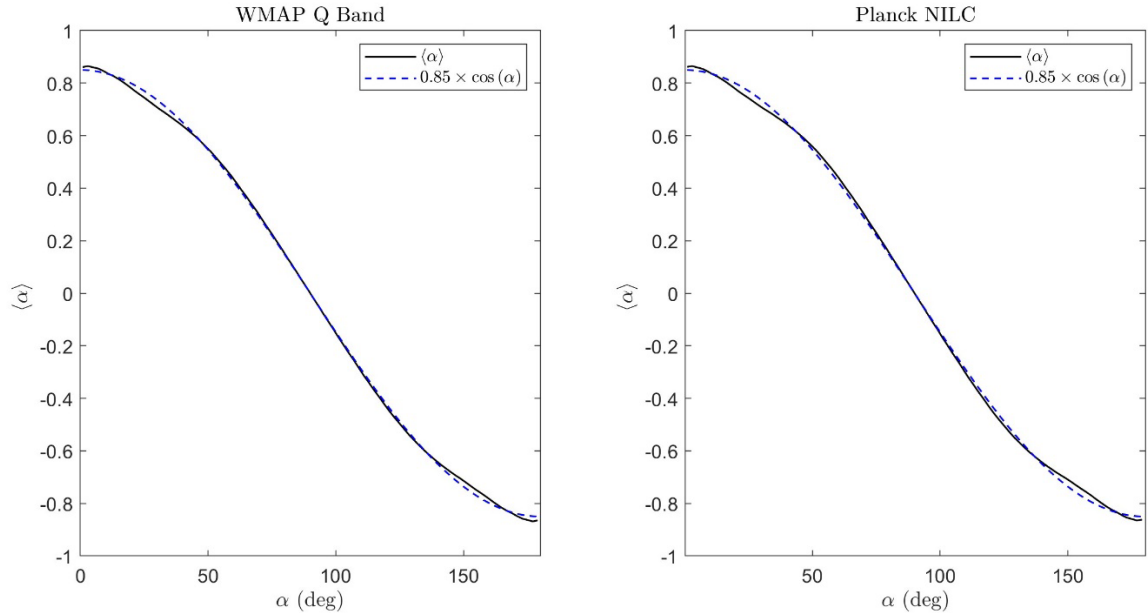

*Figure 1A:* These two panels figure shows the expected values curve $\langle \alpha \rangle \equiv \langle F(\hat{x})F(\hat{y}), \alpha \rangle$ obtained from the WMAP and Planck (left and right panels black lines, respectively) satellites' databases. As in figure 4, both curves have been calculated through a mixed probabilities method. Those new $\langle \alpha \rangle$ curves are well approached by the function $0.85\cos(\alpha)$ (dashed blue line).

Due to the significant difference respect the COBE FIRAS outcome, the four-angle set scanning was accomplished to find inequality (16) violations. In both cases several violations can be found, table 1A shows a sample of these violations.

| Satellite | $\langle\alpha_1\rangle$ | $\langle\alpha_2\rangle$ | $\langle\alpha_3\rangle$ | $\langle\alpha_4\rangle$ | $|\langle C\rangle|$ | Absolute Error |
|---|---|---|---|---|---|---|
| WMAP | 133 | 49 | 137 | 139 | 2,3972 | 0,0009 |
| WMAP | 133 | 49 | 135 | 141 | 2,3958 | 0,0009 |
| WMAP | 129 | 45 | 137 | 139 | 2,3950 | 0,0009 |
| Planck | 43 | 135 | 45 | 47 | 2,4167 | 0,0010 |
| Planck | 43 | 137 | 45 | 49 | 2,4139 | 0,0010 |
| Planck | 133 | 49 | 137 | 139 | 2,4138 | 0,0009 |

*Table 1A:* This table contains three sample sets per satellite (WMAP and Planck) of four-angle set (in deg.) each one. For each satellite, the three highest values obtained for $|\langle C\rangle|$ through the scanning procedure, are presented.



\end{document}